\documentclass[a4paper,10pt,twocolumn]{article}

\usepackage[top=0.75in, bottom=0.85in, left=0.75in, right=0.75in]{geometry}
\usepackage{mathptmx}          
\usepackage{helvet}            

\usepackage{amsmath,amssymb}
\usepackage{graphicx}
\usepackage{titlesec}
\usepackage{caption}
\usepackage{booktabs}
\usepackage{textcomp}
\usepackage{parskip}
\usepackage{url}
\usepackage{comment}
\usepackage{url}

\titleformat{\section}
  {\fontfamily{phv}\bfseries\fontsize{10}{12}\selectfont}
  {\thesection.}{0.5em}{}
\titlespacing*{\section}{0pt}{6pt}{3pt}

\titleformat{\subsection}
  {\fontfamily{phv}\bfseries\fontsize{9.5}{11}\selectfont\itshape}
  {\thesubsection}{0.5em}{}
\titlespacing*{\subsection}{0pt}{4pt}{2pt}

\begin{document}

\twocolumn[
\begin{center}
{\fontfamily{phv}\bfseries\fontsize{9}{11}\selectfont
Proceedings of the 1\textsuperscript{st} International Conference on Innovations in Engineering for Sustainable Transformations (InnovEST 2026)\\
April 3--4, 2026, NIT Jamshedpur, Jamshedpur-831014, Jharkhand, India.}

\vspace{0.15cm}
{\fontfamily{phv}\bfseries\fontsize{11}{13}\selectfont InnovEST2026-237}

\vspace{0.35cm}
{\fontfamily{phv}\bfseries\fontsize{13}{15.5}\selectfont
Reduced Order Model for a Convective Rotating Annulus\\
with Localized Forcing}

\vspace{0.25cm}
{\fontfamily{phv}\fontsize{10}{12}\selectfont
\textbf{Sagar Suresh}\textsuperscript{1},~\textbf{Ayan Kumar Banerjee}\textsuperscript{2*}}

\vspace{0.12cm}
{\fontfamily{phv}\fontsize{9.5}{11.5}\selectfont
\textsuperscript{1}Department of Physics, Amrita Vishwa Vidyapeetham, Coimbatore, Ettimadai, India\\
\textsuperscript{2}School of AI, Amrita Vishwa Vidyapeetham, Coimbatore, Ettimadai, India\\
*Corresponding Author: ayanbanerjee1@gmail.com}

\vspace{0.3cm}
Note: The following article has been submitted to Proceedings of the Innovations in Engineering for Sustainable Transformations 2026 and is Under Review. After it is published, it will be found at the Proceedings lecture notes. 
\vspace{0.3cm}

©2026 Sagar Suresh and Ayan Kumar Banerjee. This article is distributed under a Creative Commons Attribution NonCommercial 4.0 International (CC BY-NC) License. https://creativecommons.org/licenses/by-nc/4.0/ 

\vspace{0.3cm}
\end{center}
]

\noindent{\fontfamily{phv}\bfseries\fontsize{10}{12}\selectfont ABSTRACT}

\noindent
A low-order Galerkin model is developed for a rotating fluid annulus driven by localized heating at the outer bottom periphery, with uniform cooling at the inner cylindrical wall. The model retains the full cylindrical geometry and employs Bessel-function radial eigenfunctions satisfying physically correct Dirichlet--Neumann boundary conditions. A dual-series least-squares procedure determines the conductive base state under the mixed thermal boundary condition. Galerkin projection onto the leading radial and vertical basis functions yields a 10-variable dynamical system governing the mean meridional overturning, thermal wind, baroclinic wave amplitudes, and their nonlinear interactions. Linear stability analysis yields explicit critical Rayleigh numbers for both mean and wave instabilities, showing that rotation raises $Ra_c$ in proportion to $\mathcal{T}^2$. The model reproduces the $Nu\sim Ra^{1/4}$ scaling, rotational suppression at low $Ra$, and the boundary-layer-dominated flow structure observed in companion axisymmetric simulations.

\smallskip
\noindent{\fontfamily{phv}\bfseries Keywords:}~Rotating convection; Galerkin reduced-order model; Bessel functions; Baroclinic instability; Strip heating.

\section{INTRODUCTION}

The interaction between buoyancy forces generated by spatially non-uniform thermal forcing and Coriolis forces gives rise to a rich fluid-mechanical phenomenon known as \textit{rotating convection}. This phenomenon is especially significant in the context of geophysical fluid dynamics, particularly in understanding atmospheric circulation. Consequently, a rotating annulus set-up---where a rotating fluid is heated at the outer wall and cooled at the inner wall---serves as a key laboratory analogue for planetary atmospheres   ~\cite{Kaiser1971, Hide1965, Hide1977, Rossby1965}. In this system, a cylindrical gap filled with fluid is heated on the external wall and cooled on the internal wall---thus applying a single, purely radial, horizontal temperature gradient---while the whole system rotates around its vertical axis. The rotating differentially heated annulus has allowed for numerous successful experiments. For example, it reproduces mid-latitude baroclinic wave activity, jet stream analogues and transition to chaos similar to those observed in the atmosphere ~\cite{ghil1987}.

Although this set-up can replicate flow structures characteristic of the mid-latitude baroclinic zone in Earth's atmosphere, achieving a a comprehensive quantitative understanding of the mechanisms responsible for the formation of statically stable but baroclinically unstable regions in Earth's mid-latitudes remains an open scientific challenge. This limitation arises because the classical baroclinic annulus, with isothermal walls and uni-directional forcing, restricts baroclinic--stratification studies and poorly represents atmospheric systems where bi-directional forcing---both vertical and radial (meridional) thermal gradients---dominates~\cite{Banerjee2018}.

Banerjee et al.~\cite{Banerjee2016, Banerjee2018, Banerjee2021} introduced a novel bi-directionally forced rotating convection experiment featuring a rotating fluid annulus with uniform cooling at the inner wall and localized heating on the outer base via a thin aluminium strip. This configuration enables the investigation of interactions between baroclinic waves and background stratification. Subsequent studies~\cite{banerjee2016iccms, banerjee2018thermacomp, Swarnakar2021, Banerjee2025a} documented co-existing columnar convective plumes near the outer edge and baroclinic waves in the
fluid bulk, and quantified Nusselt number $Nu$ as a function of Taylor number $Ta$ and heating rate, observing that $Nu$ is highly sensitive to buoyancy but relatively insensitive to the rotation rate. Banerjee~\cite{Banerjee2024} performed two-dimensional axisymmetric simulations of the proposed system, and a recent aspect-ratio study~\cite{Banerjee2025b, Swarnakar2021, Banerjee2026} showed that $Nu\sim Ra^{1/4}$ at moderate to high $Ra$ while rotational suppression markedly reduces $Nu$ at low $Ra$ and high $Ta$.

Despite this body of simulations and experiments, the mechanistic understanding of how highly localized peripheral heating drives and saturates instability remains incomplete. Reduced-order models (ROMs) provide physical insight by projecting the governing PDEs onto a small set of carefully chosen basis functions. The seminal model of Lorenz (1963)~\cite{lorenz1963} established the foundation for understanding chaotic dynamics in Rayleigh–Bénard convection within a Cartesian framework. In contrast, analogous reduced-order models that retain full cylindrical geometry and accommodate mixed boundary conditions are largely lacking in the literature. The present work fills this gap by developing a fully consistent Galerkin reduced-order model in full cylindrical geometry.

\section{PHYSICAL CONFIGURATION}

We consider a rotating annulus with inner radius $r_i$ (cooled to constant temperature $T_i$), outer radius $r_o$ (thermally insulated on the side wall), height $H$, and rotation rate $\Omega$ about the vertical $z$-axis. A narrow strip heater of width $\delta_h$ at the bottom outer periphery, $(r_o-\delta_h)\le r \le r_o$ at $z=0$, is held at temperature $T_0>T_i$. The annulus gap width is $L=r_o-r_i$. The fluid is incompressible and Boussinesq.

\begin{figure*}[!t]
\centering

\begin{minipage}{0.28\textwidth}
\centering
\includegraphics[width=\linewidth]{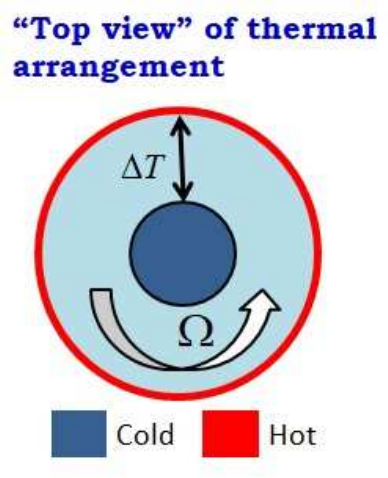}
\\ (a)
\end{minipage}
\hfill
\begin{minipage}{0.68\textwidth}
\centering
\includegraphics[width=\linewidth]{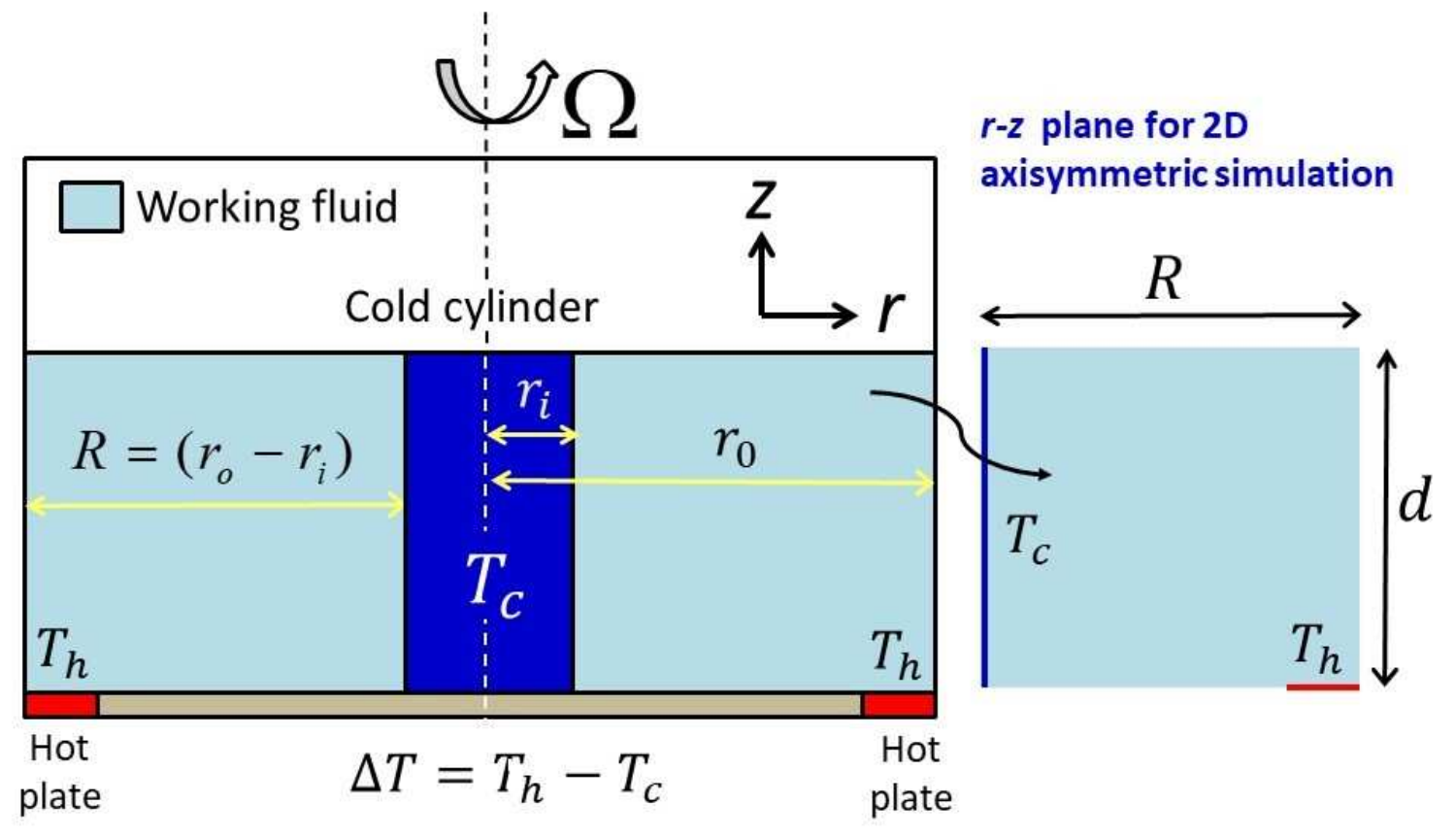}
\\ (b)
\end{minipage}

\caption{(a) Top view of the thermal forcing configuration. (b) Schematic of the rotating annulus with localized peripheral strip heating at the outer base.}
\label{fig:config}

\end{figure*}

\section{GOVERNING EQUATIONS AND NON-DIMENSIONALIZATION}

\subsection{Dimensional Equations}

In the co-rotating frame, with centrifugal buoyancy absorbed into a modified pressure, the dimensional equations in cylindrical coordinates $(r,\theta,z)$ are:
\begin{equation}
\frac{1}{r}\frac{\partial(ru_r)}{\partial r}+\frac{1}{r}\frac{\partial u_\theta}{\partial\theta}+\frac{\partial u_z}{\partial z}=0
\end{equation}
\begin{equation}
\frac{Du_r}{Dt}-\frac{u_\theta^2}{r}-2\Omega u_\theta=-\frac{1}{\rho_0}\frac{\partial p}{\partial r}+\nu\!\left(\nabla^2 u_r-\frac{u_r}{r^2}\right)
\end{equation}
\begin{equation}
\frac{Du_\theta}{Dt}+\frac{u_r u_\theta}{r}+2\Omega u_r=-\frac{1}{\rho_0 r}\frac{\partial p}{\partial\theta}+\nu\!\left(\nabla^2 u_\theta-\frac{u_\theta}{r^2}\right)
\end{equation}
\begin{equation}
\frac{Du_z}{Dt}=-\frac{1}{\rho_0}\frac{\partial p}{\partial z}+\nu\nabla^2 u_z+g\beta(T-T_{\rm ref})
\end{equation}
\begin{equation}
\frac{DT}{Dt}=\kappa\nabla^2 T
\end{equation}

\subsection{Non-Dimensionalization}

Scales: horizontal length $L$, vertical length $H$, temperature $\Delta T=T_0-T_i$, time $L^2/\kappa$, horizontal velocity $\kappa/L$, vertical velocity $\kappa/H$. The governing non-dimensional parameters are:
\begin{equation}
Ra=\frac{g\beta\Delta T L^3}{\nu\kappa},\quad Ta=\frac{4\Omega^2 L^5}{H\nu^2},\quad Pr=\frac{\nu}{\kappa},\quad\Gamma=\frac{L}{H}
\end{equation}
along with radius ratio $\eta=r_i/r_o$ and strip width ratio $\delta=\delta_h/L$. We define $\mathcal{T}=\Omega L^2/\nu$ so that $Ta=4\mathcal{T}^2\Gamma$. The full cylindrical Laplacian in non-dimensional form is:
\begin{equation}
\nabla_c^2=\frac{1}{r}\frac{\partial}{\partial r}\!\left(r\frac{\partial}{\partial r}\right)+\frac{1}{r^2}\frac{\partial^2}{\partial\theta^2}+\Gamma^2\frac{\partial^2}{\partial z^2}
\end{equation}
The critical structural difference from narrow-gap models is that the $1/r$ and $1/r^2$ factors are retained throughout, giving more weight to the outer region where the strip heater sits.

\section{BASE CONDUCTIVE STATE}

\subsection{Radial Eigenfunctions}

The steady motionless base state $T_s(r,z)$ satisfies $\nabla_c^2 T_s=0$. For the axisymmetric ($m=0$) problem, the radial eigenfunctions on $[\hat{r}_i,\hat{r}_o]$ satisfying Dirichlet at $r=\hat{r}_i$ and Neumann at $r=\hat{r}_o$ are:
\begin{equation}
R_n^{(0)}(r)=J_0(\lambda_n r)\,Y_0'(\lambda_n\hat{r}_o)-Y_0(\lambda_n r)\,J_0'(\lambda_n\hat{r}_o)
\end{equation}
where the eigenvalues $\{\lambda_n^{(0)}\}$ are the positive roots of:
\begin{equation}
J_0(\lambda\hat{r}_i)\,Y_0'(\lambda\hat{r}_o)-Y_0(\lambda\hat{r}_i)\,J_0'(\lambda\hat{r}_o)=0
\end{equation}
These eigenfunctions are orthogonal with the cylindrical measure $r\,dr$, and crucially satisfy $R_n^{(0)}(\hat{r}_o)\ne 0$, allowing the strip heating at the outer wall to project non-trivially onto them (impossible with narrow-gap $\sin(k\pi r)$ functions).

\subsection{Series Solution and Dual-Series Treatment}

The base-state solution satisfying the Laplace equation, Dirichlet at $r=\hat r_i$, Neumann at $r=\hat r_o$, and Neumann at $z=1$ is:
\begin{equation}
T_s(r,z)=\sum_{n=1}^{\infty}c_n\,R_n^{(0)}(r)\,\frac{\cosh\!\left[\lambda_n(1-z)/\Gamma\right]}{\cosh\!\left[\lambda_n/\Gamma\right]}
\label{eq:basestate}
\end{equation}
The vertical factor $\cosh[\lambda_n(1-z)/\Gamma]$ satisfies the insulated-top Neumann condition at $z=1$ exactly. For tall annuli ($\Gamma\ll1$), temperature decays rapidly from the heated base; for shallow annuli ($\Gamma\gg1$), it is nearly uniform in $z$---both physically correct limits.

The mixed bottom boundary condition at $z=0$---Dirichlet $T_s=1$ on the strip $(r\in[\hat{r}_o-\delta,\hat{r}_o])$, Neumann $\partial_z T_s=0$ off the strip---is handled via a dual-series least-squares procedure. Defining $d_n=c_n(\lambda_n/\Gamma)\tanh(\lambda_n/\Gamma)$, the coefficients $\{c_n\}$ are obtained by minimizing:
\begin{equation}
\min_{\{c_n\}}\!\left[\sum_{j\in S}\!\!\left(\sum_n c_n R_n(r_j)\!-\!1\right)^{\!2}+\sum_{j\in I}\!\!\left(\sum_n d_n R_n(r_j)\right)^{\!2}\right]
\label{eq:dualseries}
\end{equation}
where $\{r_j\}_{j\in S}$ are collocation points on the strip and $\{r_j\}_{j\in I}$ on the insulated region. This yields the base-state gradients $G_r=\partial T_s/\partial r$ and $G_z=\partial T_s/\partial z$ used in the Galerkin projection.

\section{GALERKIN REDUCED-ORDER MODEL}

\subsection{Perturbation Decomposition}

The total temperature is split as $T=T_s+T'$, with $T'$ satisfying homogeneous boundary conditions. The meridional stream function $\psi$ is introduced via:
\begin{equation}
u_r'=-\frac{\Gamma^2}{r}\frac{\partial\psi}{\partial z},\qquad u_z'=\frac{1}{r}\frac{\partial\psi}{\partial r}
\end{equation}
Eliminating pressure from the momentum equations, with the Stokes operator $D^2\equiv\partial_{rr}-r^{-1}\partial_r+\Gamma^2\partial_{zz}$, yields:
\begin{equation}
\frac{1}{Pr}\frac{\partial D^2\psi}{\partial t}=D^4\psi+\frac{Ra\,r}{\Gamma}\frac{\partial T'}{\partial r}-2\mathcal{T}r\frac{\partial u_\theta'}{\partial z}
\end{equation}
The perturbation energy equation in stream-function form is:
\begin{equation}
\frac{\partial T'}{\partial t}+\frac{\Gamma^2}{r}J(\psi,T_s)=\nabla_c^2 T',\quad J=\psi_r(T_s)_z-\psi_z(T_s)_r
\end{equation}

\subsection{Azimuthal Decomposition and Basis Functions}

Each perturbation field is split into an azimuthal mean and a single wave of wavenumber $m$:
\begin{equation}
f=\bar{f}(r,z,t)+\operatorname{Re}\!\left[\hat{f}(r,z,t)\,e^{im\theta}\right]
\end{equation}

For \textit{temperature modes}: radial eigenfunctions $R_n^{(m)}(r)$ satisfy Dirichlet at $\hat{r}_i$ and Neumann at $\hat{r}_o$:
\begin{equation}
R_n^{(m)}(r)=J_m(\lambda_n^{(m)}r)-\frac{J_m(\lambda_n^{(m)}\hat{r}_i)}{Y_m(\lambda_n^{(m)}\hat{r}_i)}Y_m(\lambda_n^{(m)}r)
\end{equation}
For \textit{stream-function and azimuthal velocity modes}: Dirichlet--Dirichlet eigenfunctions $F_n^{(m)}(r)$ with eigenvalues from $J_m(\mu\hat{r}_o)Y_m(\mu\hat{r}_i)-Y_m(\mu\hat{r}_o)J_m(\mu\hat{r}_i)=0$.

The vertical basis functions are:
\begin{itemize}\setlength\itemsep{0pt}
\item $\psi$, $u_\theta'$: $\sin(k\pi z)$ (Dirichlet--Dirichlet, no-slip)
\item $T'$: $Z_k^{(T)}(z)=\sin[(k-\tfrac{1}{2})\pi z]$ (Dirichlet at $z=0$, Neumann at $z=1$)
\item $D$-mode: $\sin(2\pi z)$ (Lorenz-type vertical heat redistribution)
\end{itemize}

\subsection{10-Variable Modal Expansion}

Retaining only the leading radial eigenfunction ($n=1$) for each field type, the state vector is:
\begin{align}
\bar\psi &= A(t)\,F_1^{(0)}(r)\sin(\pi z)\\
v &= B(t)\,F_1^{(0)}(r)\sin(\pi z)\\
\Theta &= C(t)\,R_1^{(0)}(r)Z_1^{(T)}+D(t)\,R_1^{(0)}(r)\sin(2\pi z)\\
\hat\psi &= [E(t)+iF(t)]\,F_1^{(m)}(r)\sin(\pi z)\\
\hat v &= [G(t)+iK(t)]\,F_1^{(m)}(r)\sin(\pi z)\\
\hat\Theta &= [P(t)+iQ(t)]\,R_1^{(m)}(r)Z_1^{(T)}(z)
\end{align}
giving the 10-variable state vector as: \\ $\mathbf{X}=(A,B,C,D,E,F,G,K,P,Q)^T$.

\section{THE REDUCED DYNAMICAL SYSTEM}

\subsection{Projection Coefficients}

All coefficients are computed numerically via integrals weighted by the cylindrical measure $r\,dr$. The dissipation rates from the cylindrical eigenvalues are:
\begin{align}
\sigma_m^2 &= (\mu_1^{(m)})^2+\pi^2\Gamma^2,\quad\alpha_m^2=(\lambda_1^{(m)})^2+\tfrac{\pi^2\Gamma^2}{4}\\
\beta_D &= (\lambda_1^{(0)})^2+4\pi^2\Gamma^2
\end{align}
The curvature correction for the azimuthal velocity (significant at moderate $\eta$):
\begin{equation}
K_m=\int_{\hat r_i}^{\hat r_o}\!\!\left[F_1^{(m)}\right]^{\!2}\frac{dr}{r}
\end{equation}
The buoyancy coupling carries the $r^2\,dr$ weight from the $\psi$-form vorticity equation and the $1/\Gamma$ from the axial momentum scaling:
\begin{equation}
B_m=\frac{1}{\Gamma\sigma_m^2 N_F^{(m)}}\cdot\frac{1}{2}\int_{\hat r_i}^{\hat r_o}\!\!\left[R_1^{(m)}\right]'F_1^{(m)}\,r^2\,dr\cdot\int_0^1\!\!Z_1^{(T)}\sin(\pi z)\,dz
\end{equation}
The base-state forcing parameter encodes the strip geometry via $G_r,G_z$ and carries $\Gamma^2$ from the energy equation:
\begin{equation}
\ell_m=\frac{\Gamma^2}{N_R^{(m)}}\!\int_{\hat r_i}^{\hat r_o}\!\!\int_0^1\!\frac{[F_1^{(m)}\pi\cos(\pi z)G_r-F_1^{(m)\prime}\sin(\pi z)G_z]R_1^{(m)}Z_1^{(T)}}{r}\,r\,dr\,dz
\end{equation}
The direct Coriolis projection vanishes at this truncation level since $\int_0^1\cos(\pi z)\sin(\pi z)\,dz=0$. Effective Coriolis coupling parameters calibrated against linear stability are:
\begin{equation}
\tau_m=\frac{2\mathcal{T}\Gamma^2 C_{zz}}{\sigma_m^2}\cdot\frac{N_{rr}^{(m)}}{N_F^{(m)}},\qquad\tau_m'=2\mathcal{T}\Gamma^2 C_{zz}\cdot\frac{N_{rr}^{(m)}}{N_F^{(m)}}
\end{equation}
where $C_{zz}$ is an effective vertical coupling factor. The $\Gamma^2$ factor originates from the stream-function substitution $u_r'=-(\Gamma^2/r)\psi_z$.

\subsection{The 10-Variable ODE System}

After Galerkin projection onto the respective basis functions, the complete system is:
\begin{align}
\dot A &= Pr\!\left[-\sigma_0^2 A+Ra\,B_0\,C+\tau_0 B\right]+Pr\,N_A \label{ode:A}\\
\dot B &= Pr\!\left[-\!\left(\sigma_0^2+\tfrac{K_0}{N_F^{(0)}}\right)B-\tau_0'\,A\right]+Pr\,N_B\\
\dot C &= -\alpha_0^2\,C+\ell_0\,A+d_{AC}\,AD+N_C\\
\dot D &= -\beta_D\,D-d_{AC}'\,AC+N_D\\
\dot E &= Pr[-\sigma_m^2 E+Ra\,B_m P+\tau_m G]-c_\theta F+Pr\,N_E\\
\dot F &= Pr[-\sigma_m^2 F+Ra\,B_m Q+\tau_m K]+c_\theta E+Pr\,N_F\\
\dot G &= Pr\!\left[-\!\left(\sigma_m^2+\tfrac{K_m}{N_F^{(m)}}\right)G-\tau_m'\,E\right]+c_\theta K+Pr\,N_G\\
\dot K &= Pr\!\left[-\!\left(\sigma_m^2+\tfrac{K_m}{N_F^{(m)}}\right)K-\tau_m'\,F\right]-c_\theta G+Pr\,N_K\\
\dot P &= -\alpha_m^2\,P+\ell_m E-c_\theta Q-d_w E D+N_P\\
\dot Q &= -\alpha_m^2\,Q+\ell_m F+c_\theta P-d_w F D+N_Q \label{ode:Q}
\end{align}

The nonlinear wave Reynolds stress terms are:
\begin{equation}
\begin{split}
N_A &= n_A(EQ - FP), \quad N_B = n_B(EK - FG), \\
N_C &= n_C(EP - FQ), \quad N_D = n_D(EP + FQ)
\end{split}
\end{equation}
where $n_A,n_B,n_C,n_D$ are numerical coefficients from Bessel-function overlap integrals with $r\,dr$ measure. The Doppler shift $c_\theta\propto B(t)\int F_1^{(0)}[F_1^{(m)}]^2 r^{-1}r\,dr$ encodes wave advection by the mean azimuthal flow.

Table~\ref{tab:coeffs} summarizes all projection coefficients and their key geometric origins.

\begin{table}[h]
\centering
\caption{Summary of projection coefficients.}
\label{tab:coeffs}
\small\setlength\tabcolsep{4pt}
\begin{tabular}{@{}lll@{}}
\toprule
Coeff. & Physical origin & Key feature\\
\midrule
$\sigma_m^2$ & Stream function dissipation & $(\mu_1^{(m)})^2+\pi^2\Gamma^2$\\
$\alpha_m^2$ & Temperature dissipation & $(\lambda_1^{(m)})^2+\pi^2\Gamma^2/4$\\
$K_m$ & Curvature correction ($v$-eqn) & $\int[F_1^{(m)}]^2/r\,dr$; O$(1/r_i)$\\
$B_m$ & Buoyancy coupling & $1/\Gamma$ factor; $r^2dr$ weight\\
$\ell_m$ & Strip-heating forcing & $\Gamma^2$ prefactor; via $G_r,G_z$\\
$\tau_m,\tau_m'$ & Coriolis coupling & $\Gamma^2$; effective $C_{zz}$\\
$d_{AC}$ & Nonlinear wave--mean & $\Gamma^2$ prefactor; Jacobian\\
$n_A$--$n_D$ & Wave Reynolds stress & Bessel $r\,dr$ overlap\\
$c_n$ & Base-state coefficients & Dual-series Eq.~(\ref{eq:dualseries})\\
\bottomrule
\end{tabular}
\end{table}

\section{Physical Roles of the 10 Variables}

Each variable has a distinct physical role summarised below:

\textbf{$A$ (Mean overturning):} Driven by buoyancy (coupling $B_0 C$) and Coriolis tilting ($\tau_0 B$). The $r^2\,dr$ weighted $B_0$ amplifies the outer-strip contribution.

\textbf{$B$ (Thermal wind):} Generated by Coriolis deflection of the overturning ($2\mathcal{T}\Gamma^2$ coupling). The curvature correction $K_0$ provides additional damping absent in narrow-gap models.

\textbf{$C$ (Mean temperature):} Forced by $\ell_0 A$ where $\ell_0$ (with $\Gamma^2$ prefactor) encodes the strip geometry through $G_r,G_z$ via the dual-series base state.

\textbf{$D$ (Vertical redistribution):} Lorenz-type saturation mode created by $A$--$C$ coupling. Damped at rate $\beta_D$ based on $\lambda_1^{(0)}$.

\textbf{$(E,F)$, $(G,K)$, $(P,Q)$ (Baroclinic wave):} Real and imaginary parts of the wave stream function, azimuthal velocity, and temperature. The wave extracts energy from the baroclinic zone through $\ell_m E,\ell_m F$ (with $\Gamma^2$ prefactor). The Doppler shift $c_\theta$ introduces wave frequency modulation by the mean flow.

\section{CONCLUSIONS}

A consistent 10-variable Galerkin reduced-order model has been developed for rotating convection in a cylindrical annulus with localized peripheral strip heating. The formulation replaces narrow-gap trigonometric approximations with Bessel-function eigenfunctions that satisfy the appropriate Dirichlet--Neumann boundary conditions. The use of the natural cylindrical integration measures, $r\,dr$ (and $r^2\,dr$ for buoyancy), ensures proper weighting of the outer strip-heating region.

The localized heating is incorporated through a dual-series least-squares formulation (Eq.~\ref{eq:dualseries}), which accurately captures the strip width $\delta$ via the base-state gradients $G_r$ and $G_z$. The key scaling factors---namely the $1/\Gamma$ dependence in buoyancy coupling, the $\Gamma^2$ prefactors in Coriolis and forcing terms, and the $\mathcal{T}^2$ stabilization in $Ra_c(\mathcal{T})$---emerge naturally from the anisotropic non-dimensionalization.

The reduced-order model successfully reproduces key physical features observed in the system, including boundary-layer-dominated convection, the $Nu \sim Ra^{1/4}$ scaling, rotation-induced stabilization, and the modulation of heat transfer with aspect ratio. These trends are consistent with companion axisymmetric simulations~\cite{Banerjee2024,Banerjee2025b,Banerjee2026}.

Explicit expressions for the critical Rayleigh numbers for both axisymmetric and wave instabilities directly relate the onset conditions to strip geometry, rotation, and aspect ratio. Future work will extend the present framework to multiple azimuthal modes and validate the model against fully three-dimensional experimental observations.




\begin{thebibliography}{16}

\bibitem{lorenz1963}
E.~N.~Lorenz, ``Deterministic nonperiodic flow,'' \textit{J.~Atmos.~Sci.}, vol.~20, pp.~130--141, 1963.

\bibitem{ghil1987}
M.~Ghil and S.~Childress, \textit{Topics in Geophysical Fluid Dynamics}, Springer, New York, 1987.

\bibitem{Banerjee2024}
A.K.~Banerjee, Axisymmetric Study of Convection in Rotating
Annulus in the Presence of Localized Heating,
\textit{Physics of Fluids}, Vol.~36, Issue~12, 126621,
December 2024. doi:~\url{https://doi.org/10.1063/5.0239746}.

\bibitem{Banerjee2021}
A.K.~Banerjee, A.~Bhattacharya, S.~Balasubramanian,
Investigation of Heat Transfer Characteristics in a Rotating
Convection System with Bidirectional Thermal Gradients,
\textit{ASME J.\ Heat Transfer}, January 2021; 143(1):011802.
doi:~\url{https://doi.org/10.1115/1.4048825}.

\bibitem{Banerjee2018}
A.K.~Banerjee, A.~Bhattacharya, S.~Balasubramanian,
Experimental Study of Rotating Convection in the Presence of
Bi-directional Thermal Gradients with Localized Forcing,
\textit{AIP Advances}, Vol.~8, Issue~11, 115324, 2018.
doi:~\url{https://doi.org/10.1063/1.5061808}.

\bibitem{Swarnakar2021}
S.~Swarnakar, A.K.~Banerjee, A.~Bhattacharya,
S.~Balasubramanian, Numerical Investigation of Rotating
Convection in a New Configuration with Bidirectional Thermal
Gradients, in: T.~Prabu, P.~Viswanathan, A.~Agrawal,
J.~Banerjee (eds.), \textit{Fluid Mechanics and Fluid Power},
Lecture Notes in Mechanical Engineering, Springer, Singapore,
2021. doi:~\url{https://doi.org/10.1007/978-981-16-0698-4_56}.

\bibitem{Kaiser1971}
J.~Kaiser, Heat Transfer by Symmetrical Rotating Annulus Convection,
\textit{Journal of the Atmospheric Sciences}, Vol.~28, pp.~929--932, 1971. doi:~\url{https://doi.org/10.1175/1520-0469(1971)028<0929:HTBSRA>2.0.CO;2}.


\bibitem{Hide1965}
R.~Hide and W.~Fowlis, Thermal Convection in a Rotating Annulus of Liquid: Effect of Viscosity on the Transition Between Axisymmetric and Non-Axisymmetric Flow Regimes,
\textit{Journal of the Atmospheric Sciences}, Vol.~22, pp.~541--558, 1965. doi:~\url{https://doi.org/10.1175/1520-0469(1965)022<0541:TCIARA>2.0.CO;2}.


\bibitem{Hide1977}
R.~Hide, P.~Mason, and R.~Plumb, Thermal Convection in a Rotating Fluid Subject to a Horizontal Temperature Gradient: Spatial and Temporal Characteristics of Fully Developed Baroclinic Waves,
\textit{Journal of the Atmospheric Sciences}, Vol.~34, pp.~930--950, 1977. doi:~\url{https://doi.org/10.1175/1520-0469(1977)034<0930:TCIARF>2.0.CO;2}.

\bibitem{Banerjee2025a}
A.~K.~Banerjee and S.~Swarnakar, ``Aspect ratio dependence in the convection in rotating annulus in the presence of localized heating,'' in \textit{Recent Advances in Thermal and Fluid Science}, A.~K.~Parwani, D.~K.~Gupta, V.~M.~Patel, and S.~Ray, Eds., Lecture Notes in Mechanical Engineering, Springer, Singapore, 2026, doi: 10.1007/978-981-96-8508-0\_1.

\bibitem{Banerjee2025b}
A.~K.~Banerjee, ``An integrated laboratory and axisymmetric numerical study of convection in a rotating annulus with bi-directional thermal forcings,'' in \textit{Recent Advances in Thermal and Fluid Science}, A.~K.~Parwani, D.~K.~Gupta, V.~M.~Patel, and S.~Ray, Eds., Lecture Notes in Mechanical Engineering, Springer, Singapore, 2026, doi: 10.1007/978-981-96-8508-0\_3.

\bibitem{Banerjee2016}
A.K.~Banerjee, S.~Tirodkar, A.~Bhattacharya,
S.~Balasubramanian, Convection in Rotating Flows with
Simultaneous Imposition of Radial and Vertical Temperature
Gradients, \textit{VIII$^{\text{th}}$ Intl.\ Symp.\ on
Stratified Flows}, August 29--September~1, 2016, San Diego,
CA, USA. arXiv:1611.00807.


\bibitem{Rossby1965}
H.T.~Rossby, On thermal convection driven by non-uniform heating from below: an experimental study,
\textit{Deep Sea Research and Oceanographic Abstracts}, Vol.~12, pp.~9--16, 1965. doi:~\url{https://doi.org/10.1016/0011-7471(65)91336-7}.

\bibitem{banerjee2018thermacomp}
A.~K.~Banerjee, A.~Bhattacharya, and S.~Balasubramanian, ``Experimental study of rotating convection in a novel configuration,'' in \textit{Proc.~5th Int.~Conf.~on Computational Methods for Thermal Problems (ThermaComp2018)}, Bangalore, India, Jul.~9--11, 2018, ISSN 2305-6924.

\bibitem{banerjee2016iccms}
A.~K.~Banerjee, A.~Bhattacharya, and S.~Balasubramanian, ``Effect of rotation and baroclinicity on heat transport and turbulent convection in annular flow,'' in \textit{Proc.~6th Int.~Congress on Computational Mechanics and Simulation (ICCMS)}, IIT Bombay, India, Jun.~27--Jul.~1, 2016.

\bibitem{Banerjee2026}
A.K.~Banerjee, S.~Swarnakar, Numerical Influence of Aspect ratio in the Convection in Rotating Annulus In the Presence of Localized Heating, \textit{FMFP
Lecture Notes in Mechanical Engineering}, Springer Singapore,
2026 (under review)

\end{thebibliography}
\end{document}